\documentclass[aps,prb,floatfix,twocolumn,superscriptaddress,showpacs,epsf]{revtex4-2}

\usepackage[USenglish]{babel}
\usepackage{amsmath,amssymb,amsfonts}
\usepackage{epstopdf}
\usepackage{epsfig} 
\usepackage{graphicx}
\usepackage{bm}
\usepackage{subfigure}
\usepackage{import}
\usepackage{color}
\usepackage{url}
\usepackage{overpic}
\usepackage[final]{changes}

\allowdisplaybreaks

\usepackage[breaklinks=true,colorlinks,citecolor=blue,linkcolor=blue,urlcolor=blue]{hyperref}

\makeatletter
\renewcommand\tableofcontents{\@starttoc{toc}}
\makeatother
\def\bcen{\begin{center}}
\def\ecen{\end{center}}

\def\a{\alpha}       \def\b{\beta}   \def\g{\gamma}   \def\d{\delta} 
          
       \def\l{\lambda}

       \def\D{\Delta}

  \def\ie{\mbox{\it i.e.\ }}
\def\=={\equiv}

\def\qed{\raise1pt\hbox{\vrule height5pt width5pt depth0pt}}

\def\cG0{{\cal G}_0} 
\def\cG{{\cal G}}

\def\ie{\hbox{\it i.e.\ }} 
\def\ie{\mbox{\it i.e.\ }} \def\=={\equiv}

 \def\ep0{\epsilon_{p}} \def\ed0{\epsilon_{f}}

\def\be{\begin{equation}}
\def\ee{\end{equation}}

\def\cc{c^{\dagger}}
\def\ca{c^{\phantom{\dagger}}}

\newcommand{\ket}[1]{|{#1}\rangle}

\newcommand{\braket}[3]{\langle{#1}| {#2} |{#3} \rangle}

\newcommand{\bd}[1]{\mathbf{#1}}
\newcommand{\bs}[1]{\boldsymbol{#1}}

\newcommand{\quave}[1]{\langle{#1}\rangle}

\usepackage{fancyhdr}
\begin{document}

\author{Giacomo Mazza}
\email{giacomo.mazza@unipi.it}
\affiliation{Dipartimento di Fisica dell'Universit\`a di Pisa, Largo Bruno Pontecorvo 3, I-56127 Pisa, Italy}
\affiliation{Department of Quantum Matter Physics, University of Geneva, Quai Ernest-Ansermet 24, 1211 Geneva, Switzerland}
\author{Marco Polini}
\affiliation{Dipartimento di Fisica dell'Universit\`a di Pisa, Largo Bruno Pontecorvo 3, I-56127 Pisa, Italy}
\affiliation{Istituto Italiano di Tecnologia, Graphene Labs, Via Morego 30, I-16163 Genova, Italy}
\affiliation{ICFO-Institut de Ci\`{e}ncies Fot\`{o}niques, The Barcelona Institute of Science and Technology, Av. Carl Friedrich Gauss 3, 08860 Castelldefels (Barcelona),~Spain}

\title{Hidden excitonic quantum states with broken time-reversal symmetry}
\begin{abstract}

The spontaneous breaking of time-reversal symmetry due to purely-orbital mechanisms (i.e.~not involving spin degrees of freedom) 
yields extremely exotic phases of matter such as 
Chern insulators and chiral superconductors. 
In this Letter, we show that excitonic insulators,
by exploiting the transition from the excitonic ground state to 
a purely-orbital time reversal symmetry broken hidden state, 
can realize another notable example of this class.
The transition to the hidden state is controlled by engineered 
geometrical constraints which enable the coupling between the excitonic order 
parameter and the free-space electromagnetic field.
These results pave the way towards exotic orbital magnetic order 
in quantum materials and are also relevant for disentangling excitonic 
phase transitions from trivial structural ones.
\end{abstract}

\maketitle

\section{Introduction}
The physical properties of a condensed matter
system are often direct manifestations 
of well defined processes of spontaneous symmetry breaking.
Examples range from the breaking of translation 
symmetry in the formation of crystalline solids, 
to the breaking of time-reversal symmetry in 
magnetic systems and particle conservation 
in superconductors~\cite{anderson_basic_notion}.
In some cases, however, 
the nature of symmetry breaking alone may not be enough to fully 
determine the physical properties of the system. 
This can happen in the case of excitonic phase transitions~\cite{excitonic_insulators}
in which electronic states belonging to valence 
and conduction bands spontaneously hybridize due to the 
Coulomb interaction between negatively charged electrons and 
positively charged holes.

The phenomenon of exciton condensation is well studied 
in heterostructures with spatially separated electrons and 
holes~\cite{Eisenstein_ARCMP_2014,Eisenstein_Nature_2004,mac_donald_nat_phys,
Liu_quantum_hall_drag_NatPhys2017,Li_dblg_nat_phys_2017,Burg_prl_2018}.
Nonetheless, the formation of a so-called excitonic insulator 
state in crystalline solids remains a debated question~\cite{kogar_TiSe2,exp_ei_TiSe2,exp_ei_Ta2NiSe5,
ataei_MoS2,sun_cobden_excitonic_WTe2,varsano_mos2}.
The main reason is that the excitonic instability 
can generically lead to different physical 
manifestations depending on the degrees of freedom 
it couples with.
For example, the excitonic instability may 
induce a distortion of the charge density which,
due to the coupling  with the ionic degrees of freedom, 
makes the excitonic state practically indistinguishable 
from a structural phase transition~\cite{giacomo_TNS,
 kaneko_ortho_to_mono,
 watson_tns2020,lukas_TNS,
subedi_TNS,Kim2021_raman_TNS,
Ye2021_raman_TNS}.

\begin{figure}[t]
\includegraphics[width=0.95\linewidth]{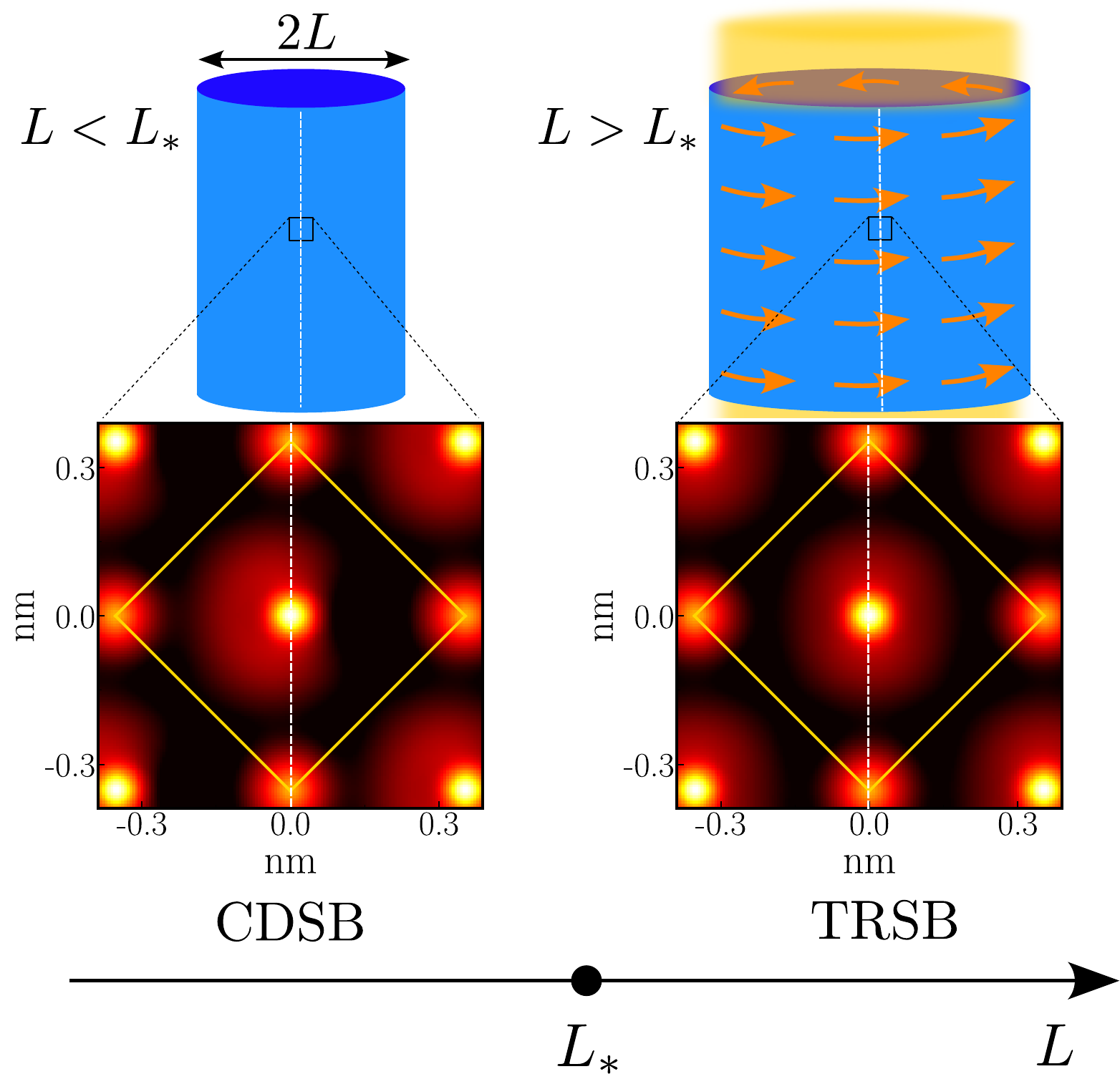}
\caption{Schematic phase diagram of the excitonic  
ground state as a function of the cylinder radius $L$. 
The color plots show the charge density 
distributions in the unit cells (indicated by yellow squares) 
for $L<L_\ast$ and $L>L_\ast$. 
The white dashed lines highlight the reflection symmetry broken 
by the excitonic instability. 
In the CDSB state the charge density is not invariant under 
the reflection symmetry. 
The TRSB state displays a solenoidal persistent current being 
characterised by a symmetric charge density.
The yellow shading is a pictorial representation of the 
self-generated flux that sustains the persistent current.}
\label{fig:fig0}
\end{figure}

In this Letter, we exploit a dichotomic 
manifestation of the symmetry breaking underlying an excitonic phase transition
to show the stabilization of a time-reversal symmetry broken (TRSB) 
hidden quantum state in a two-dimensional (2D) material 
by means of engineered geometrical constraints.
Fig.~\ref{fig:fig0} summarizes the main 
result of this Letter: The ground state of a  
2D material  harboring an excitonic phase transition, in a cylinder geometry, 
can be transformed from a charge-density symmetry broken (CDSB) 
state to a TRSB one {(which hosts persistent orbital currents)}
by controlling the cylinder radius.

The change in the ground state nature is driven by 
a self-generated flux which sustains the equilibrium orbital currents, 
and lowers the energy of the hidden excitonic TRSB state with respect to the 
energy of the CDSB ground state.
We predict the transition to occur for cylinders 
with radius larger than a critical value, which can be further 
controlled by an applied external flux.
%
%

\section{Model and purely electronic instability.} 
\begin{figure}
\begin{overpic}[width=\linewidth]{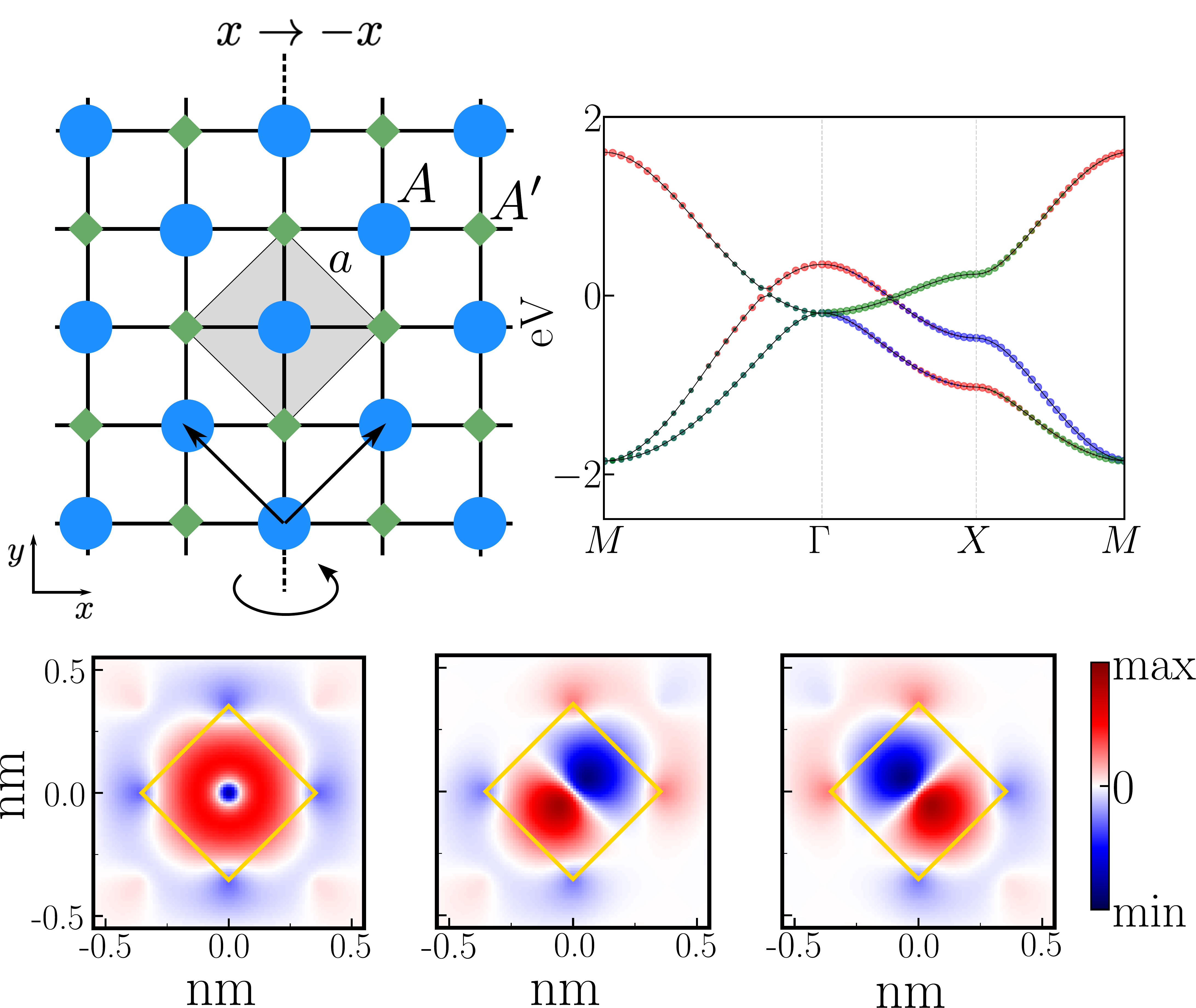}
\put(0,80) {(a)}
\put(50,80) {(b)}
\put(0,30) {(c)}
\end{overpic}
\caption{(a) 2D crystal structure. {Big {(blue)} circles and 
small {(green)} diamonds represent the positions of the two atomic-like 
potential wells, $A$  and $A'$, respectively.} 
{The shaded area denotes the unit cell, while $a=0.5~\mathrm{nm}$ indicates the lattice parameter.}
The dashed line represents the reflection 
symmetry broken by the excitonic instability 
in the channel $\D_{+}=-i\D_{-}=\D$.
The arrows indicate the primitive vectors of the Bravais lattice.
The curved arrow indicates the wrapping 
around the $y-$axis discussed in this work.
(b) Fat band representation of the low-energy 
band structure. {The band structure is obtained by setting (see main text): 
$\g_{A} = 1.52~\mbox{eV}~\mbox{nm} $, 
$\g_{A'} = 0.38~\mbox{eV}~\mbox{nm}$,
$\xi_A = \xi_{A'} = 1~\mbox{nm}$, and 
$\eta_{A}=\eta_{A'} = 10^{-3}~\mbox{nm}$. Red, green, and blue projections 
correspond, respectively, to the $s$, $p_+$, and $p_-$ Wannier 
orbitals.} 
(c) Real-space maps of the $s$, $p_+$, and $p_-$ Wannier orbitals (from left to right). \label{fig:fig1}}
\end{figure}

{We consider a system of interacting 
electrons subject to a 2D potential with the 
periodicity of a square lattice,} 
$V({\bm x}) = V({\bm x}+ {\bm R})$.  The crystalline potential 
$V({\bm x}) = \sum_{{\bm R}, a} v_{a} ({\bm x} - {\bm R})$ 
originates from atomic-like potential wells $v_{a}({\bm x} - {\bm R})$ centered
at the position of an atom of type $a$ in the unit cell ${\bm R}$.
Specifically, we consider two atoms $a = A, A^\prime$ per unit cell, arranged as in Fig.~\ref{fig:fig1}(a).
We choose atomic potentials of the Yukawa type, 
i.e.~$v_{a}(\bs{x}) = - \g_{a} e^{-|\bs{x}|/\xi_a}/(|\bs{x}| + \eta_{a})$, where $\eta_{a}$ is a short-distance cut-off. 
We solve Bloch eigenvalue problem and 
adjust the parameters $\g_a,\xi_a,$ and $\eta_a$ in order 
to obtain a band structure in which bands originating from 
atomic-like orbitals of different symmetry overlap at the 
Fermi level. 
In this work we consider spinless fermions and an excitonic instability that occurs only in the 
spin singlet channel~\cite{amaricci_BHZ}. 

By assuming four electrons per unit cell, the 
have a bandwidth $W~\sim 4~\mbox{eV}$
and mainly originate from the $2s$ orbital and 
$1 p_{\pm} = 1p_{x} \pm i 1p_y$ orbitals of the central atom 
(Fig.~\ref{fig:fig1}(b)).
The three low-energy bands are occupied by two electrons, 
whereas two additional electrons occupy two core bands 
(not shown) originating, respectively, from the $1s$ orbitals 
of the central and corner atoms.
We derive localized Wannier wavefunctions using the 
projection method~\cite{Marzari_RMP}
and write the single-particle band Hamiltonian in the Wannier basis as
${\cal H}_0 = \sum_{{\bm R}, {\bm  R}^\prime} \sum_{\a, \b = s,p_{\pm}} 
h^{\a \b}_{{\bm R}, {\bm R}^\prime} \cc_{{\bm R}, \a} \ca_{{\bm R}^\prime, \b}$.
The Wannier wavefunctions are mainly localized on the central atom, 
with weaker amplitude on the corner atoms---see Fig.~\ref{fig:fig1}(c).
The $s$-like wave-function is even under reflection with respect to  
the two axes $y=\pm x$, whereas the $p_{\pm}$-like ones
are odd/even under reflections with respect 
to the $y=\pm x / \mp x$ axes, respectively.

Due to the above symmetries, the local matrix elements 
of the single-particle Hamiltonian ${\cal H}_0$ between $s$- and $p$-like orbitals identically vanish, 
i.e.~$h_{{\bm R}, {\bm R}}^{sp_{\pm}} =0$, 
implying a vanishing local hybridization between $s$ and $p_{\pm}$ 
Wannier wave-functions:
\begin{equation}
\D^{(0)}_{\pm} \equiv \quave{\cc_{{\bm R}, s} \ca_{{\bm R}, p_{\pm}}} = 0~.
\label{eq: symmetry}
\end{equation}

The hybridizations in Eq.~(\ref{eq: symmetry}) play the role of 
order parameters for the breaking of the symmetries of the square lattice
due to the excitonic instability. With this target in mind, 
we add to the bare band Hamiltonian ${\cal H}_0$ two electron-electron interaction contributions: 
i) a purely local density-density interaction $U$ between $p_+$ and $p_-$ Wannier orbitals, 
and 
ii) a density-density  interaction $V$ between $s$ and $p_{\pm}$ orbitals. 
The full Hamiltonian therefore reads:
\begin{equation}
{\cal H} = {\cal H}_0 + U \sum_{\bm R} n_{{\bm R}, p_{+}} n_{{\bm R}, p_{-}}
+ V \sum_{{\bm R}, \a = \pm} n_{{\bm R}, p_{\a}} n_{{\bm R}, s}~.  
\end{equation}
For the sake of concreteness, we fix $U=2.5~\mathrm{eV}$ and 
$V=2.0~\mathrm{eV}$, of the same order of magnitude of the 
bandwidth $W$.
 
Seeking for the excitonic instability, we introduce a family of 
variational wave-functions $\ket{\Psi_{\D}}$, 
corresponding to BCS wave functions with 
pairing in the particle-hole channel, and allow for a non-vanishing order parameter
\begin{equation}\label{eq:order-parameter}
\D_{\pm} \equiv \braket{\Psi_{\D}}{\cc_{{\bm R}, p_{\a}} \ca_{{\bm R}, s}}{\Psi_{\D}} \neq 0~.
\end{equation}

By introducing Lagrange multipliers $\l_{\pm}$, 
we compute the variational energy as a function of $\D_{\pm}$:
\begin{equation}
\begin{split}
E_{\rm var}&[\D_\pm,\l_\pm] =  \braket{\Psi_{\D}}{\cal H}{\Psi_{\D}}
+  
\\ & \sum_{{\bm R}, \a=\pm}
\l_{\a} \left( 
\braket{\Psi_{\D}}{\cc_{{\bm R}, p_{\a}} \ca_{{\bm R}, s}}{\Psi_{\D}} 
- \D_{\a} \right) + {\rm c.c.}~.
\end{split}
\label{eq:energy_functional_electronic}
\end{equation}
For a given $\D_{\pm}$, we  self-consistently optimize the variational wave-function with respect to 
the Lagrange multipliers and all the expectation values 
$\D_{{\bm k}}^{\a ,\b} = \braket{\Psi_{\D}}{\cc_{{\bm k}, \a} \ca_{{\bm k}, \b}}{\Psi_{\D}}$. 
We restrict our analysis to the 
channel $\D_{+} = -i \D_{-}$, 
\added{which we found as the most stable channel for symmetry breaking in the model.}
\replaced{Such combination of order parameters }{which }corresponds 
to the breaking of reflection symmetry $x \to -x$ (see Fig.~\ref{fig:fig1}).
\replaced{As a result,}{With this choice,} the energy functional in 
Eq.~(\ref{eq:energy_functional_electronic}) reduces to a function 
of  a single order parameter $\D = |\D| e^{i \varphi} \equiv \D_{+}$.
\added{We notice that, as a consequence of the $s-p_{\pm}$ 
hybridizations, the symmetry broken state automatically 
displays a non-zero $p_{+}-p_{-}$ hybridization }
$\D_{pp} \equiv \braket{\Psi_{\D}}{\cc_{p_+} \ca_{p_-}}{\Psi_{\D}} \neq 0$ 
\added{due to broken $4-$fold rotation symmetry of the square lattice. }

The nature of the excitonic states issuing from the breaking of the 
$x\to-x$ symmetry depends on the order parameter phase 
$\varphi$~\cite{kopaev_current,volkov_superdiamagnetism}.
This is readily understood from simple arguments of quantum mechanics.
Let us consider two generic real wavefunctions 
$\Psi_{\rm e}(x,y) = \Psi_{\rm e}(-x,y)$ and 
$\Psi_{\rm o}(x,y)= - \Psi_{\rm o}(- x,y)$, which 
are even and odd, respectively, under the 
$x\to-x$ reflection symmetry. 
The excitonic order parameter corresponds 
to  an hybridized state $\Psi({\bm x}) \sim \Psi_{\rm e}({\bm x}) 
+ |\D| e^{i \varphi} \Psi_{\rm o}({\bm x})$.
By computing the charge $\rho({\bm x}) \equiv \Psi^*({\bm x}) \Psi({\bm x})$ 
and momentum ${\bm p}({\bm x}) \equiv - \frac{i}{2} \hbar \Psi^*({\bm x}) \nabla \Psi({\bm x}) + c.c.$ densities, it is straightforward to see that for $\varphi=0,\pi$ 
$\Psi({\bm x})$ corresponds to a CDSB state 
with $\rho(x,y) \neq \rho(-x,y)$, and which displays 
time-reversal symmetry, ${\bm p}({\bm x}) = 0$. 
On the contrary, for  
$\varphi = \pm \frac{\pi}{2} $, it corresponds to a TRSB state
with ${\bm p}(-x,y) \neq - {\bm p}(x,y)$, and a symmetric 
charge density $\rho({\bm x}) = \rho(-{\bm x})$.
For a generic $\varphi \neq 0,\pi,\pm \frac{\pi}{2} $, 
$\Psi({\bm x})$ describes at the same time a CDSB and TRSB state.

\begin{figure}
\begin{overpic}[width=\linewidth]{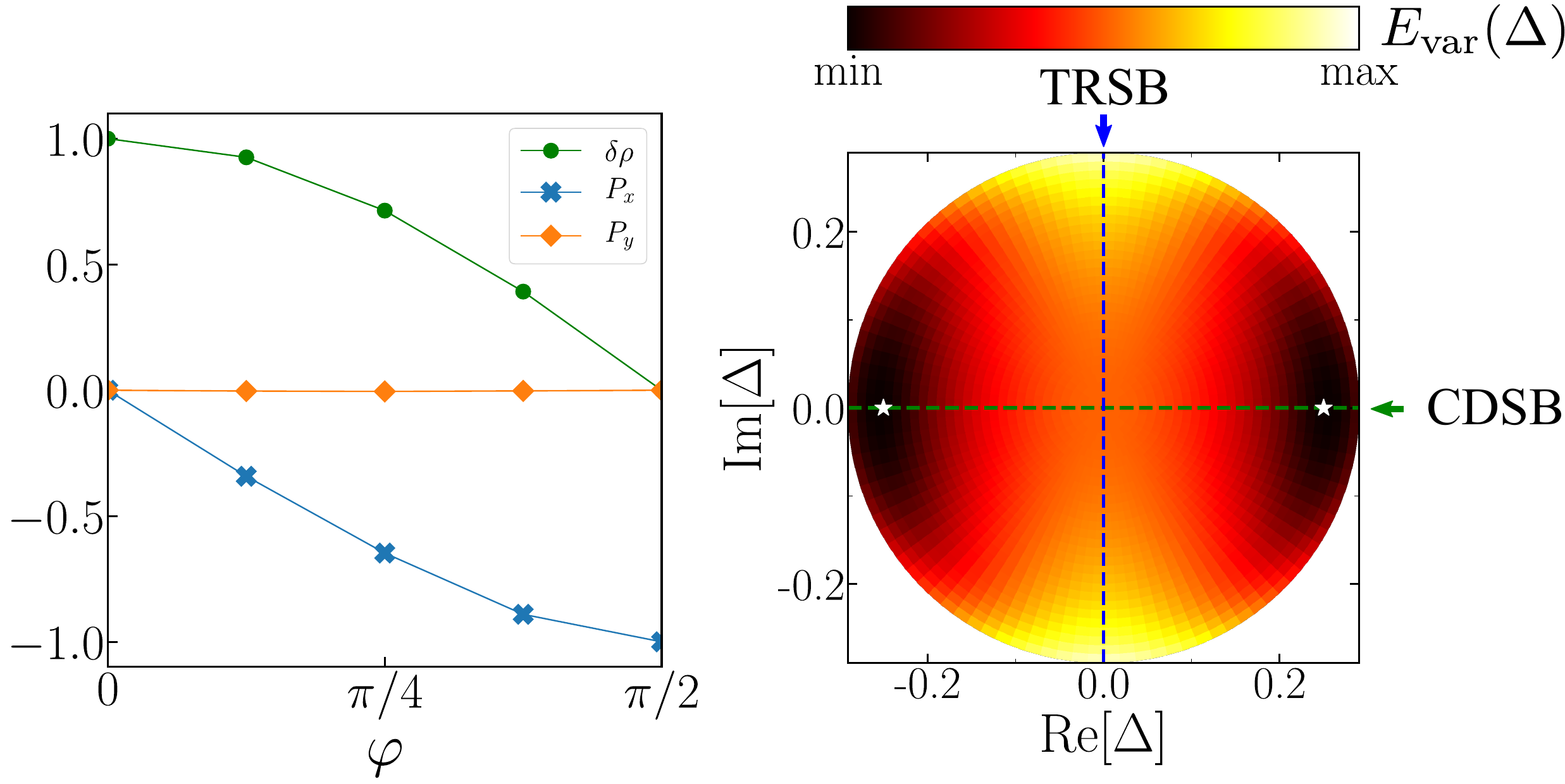}
\put(0,47.5){(a)}
\put(47,47.5){(b)}
\end{overpic}
\caption{ (a) 
Quantifiers $\delta \rho$ and ${\bm P}$ (see main text) of the CDSB and TRSB character of a generic state $|\Psi_\D\rangle$ at a fixed value of $|\D|=0.25$ and 
varying phase $\varphi$. $\delta \rho$ 
is measured with respect to its $\varphi=0$ value, 
whereas $P_{x/y}$ components are measured 
with respect to their $\varphi = \pi/2$ values. 
(b) 
Variational energy for the pure electronic problem
as a function of  $\mathrm{Re}(\D)$ and $\mathrm{Im}(\D)$. 
The white stars indicate the two global 
minima while the dashed horizontal (vertical) line
highlights the purely CDSB (TRSB) state.}
\label{fig:fig2}
\end{figure}

We explicitly show the CDSB and TRSB 
characters of the variational states $|\Psi_\Delta\rangle$ by computing 
the quantities $\d \rho \equiv \int_{x<0} \rho_\D({\bm x}) d^2 {\bm x} - 
\int_{x>0} \rho_\D({\bm x})d^2 {\bm x} $
and $ {\bm P} \equiv \int_{x<0} {\bm p}_\D({\bm x}) d^2 {\bm x} +
\int_{x>0} {\bm p}_\D({\bm x})d^2 {\bm x}$
for fixed order parameter modulus $|\D|$ and 
varying phase $\varphi$.
``$\int_{x</>0}$'' means integration
over the two portions of the unit cells separated by the $x=0$ 
axis, and $\rho_\D({\bm x})$ and ${\bm p}_\D({\bm x})$ 
are density and momentum densities computed on the variational
state $\ket{\Psi_\D}$ by expanding the fermionic fields 
over the complete set of Bloch 
wave-functions 
$\Psi({\bm x}) = \sum_{{\bm k}, n} \psi_{{\bm k}, n}({\bm x}) \ca_{{\bm k}, n}$. 
For $\varphi = 0$, we find $\delta \rho \neq 0$, 
and $P_x=P_y=0$ highlighting the pure CDSB nature 
of the excitonic state---see Fig.~\ref{fig:fig2}(a).
Increasing $\varphi$, the CDSB contribution 
$\delta \rho$ decreases and the excitonic states acquire 
TRSB character, as highlighted by ${\bm P} \neq 0$.
Eventually, the CDSB vanishes for $\varphi = \pi/2$, 
where the TRSB contribution ${\bm P}$ is maximum.  
We notice that the TRSB excitonic state corresponds to a 
net momentum through the unit cell along the $x$ direction.
On the contrary, the reflection symmetry $y \to -y$ 
is preserved and the net momentum along the $y$ direction is identically zero.

In Fig.~\ref{fig:fig2}(b) we plot the variational energy 
as a function of the real and imaginary parts of the 
order parameter $\D$. 
The energy surface displays two equivalent global 
minima at $|\D| \neq 0$ which highlight the spontaneous  symmetry 
breaking and reflect the $Z_2$ nature of the 
broken reflection symmetry $x \to -x$.
The two minima corresponds to real order 
parameters $\varphi=0,\pi$, 
thus showing that, for the purely electronic instability, 
the excitonic ground state corresponds to the CDSB state.

\section{TRSB Excitonic State.} 
All the TRSB states obtained by the variational 
optimization of Eq.~(\ref{eq:energy_functional_electronic})
are unstable hidden states with energy higher than the CDSB states 
at $\varphi=0,\pi$.
This result is in agreement with the no-go theorems 
that forbid a ground state with a non-vanishing {\it net} current.
The argument, 
originally attributed to Bloch~\cite{bloch_by_bohm}, 
follows from the observation that the energy of 
a state with net current can be made arbitrary small 
by an unphysical constant vector potential {(i.e.~unphysical since it can be gauged away)}.

We now show that, indeed, 
the TRSB excitonic state can be stabilized  by: 
i) changing the topology of the system and 
ii) allowing the TRSB state to act as a source of 
a physical magnetic flux.
We start by wrapping the 2D system on a cylinder 
of radius $L$ around the $y-$axis{---see Fig.~\ref{fig:fig1}(a)}.
In this cylindrical geometry, the 
current of the TRSB state is purely solenoidal. The net current 
vanishes and the state couples only to physical fields that 
cannot be gauged away.
Moreover, in this geometry the TRSB state can source 
a magnetic flux which is naturally expected to lead to an energy lowering.
Indeed, the fact that, in nature, currents act as sources of a magnetic flux
simply implies that the energy of the TRSB state including the 
flux sourced by the TRSB state itself should have energy lower than 
the energy of the TRSB state without self-generated flux. 
In the following, we show that, in this geometry, 
such  an energy lowering is controlled by the cylinder radius.


%
\begin{figure} 
\begin{overpic}[width=\linewidth]{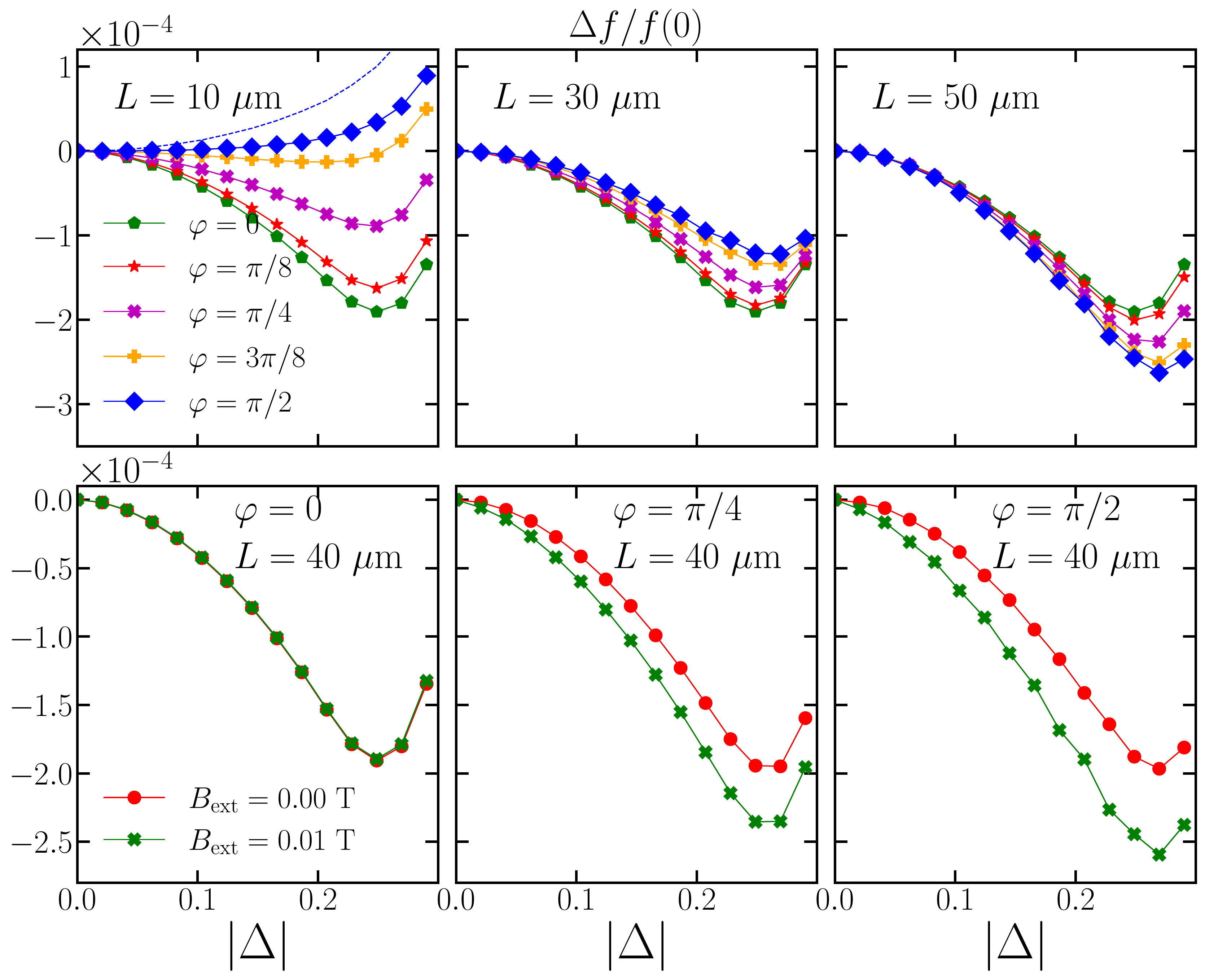}
\put(0,77){(a)}
\put(0,42){(b)}
\end{overpic}
\caption{(a) Volume energy density $f(\D)$ as 
a function of $|\D|$,
for different values of $\varphi$---from 
the pure CDSB (green pentagons) state at $\varphi=0$ to the pure TRSB (blue diamonds) state at $\varphi=\pi/2$---and different values of the cylinder radius $L$. All the energies are measured 
with respect to the energy $f(0)$ in the 
$\Delta=0$ symmetric  case---$\Delta f \equiv f(\D)-f(0)$---and in units of $f(0)$. 
In the $L=10~\mu \mbox{m}$ panel the dashed line represents the energy density 
for the $\pi/2$ state for the purely electronic problem in the absence of self-generated 
flux. (b)
Volume energy density of the pure CDSB state (left) and the pure 
TRSB state (right) as a function of an applied external flux and $L=40~\mu {\rm m}$.}
\label{fig:fig3}
\end{figure}
%


We introduce a new energy functional ${\cal E}[\D, {\bm A},\l]$ which 
depends on the excitonic order parameter $\D$, the associated Lagrange parameter $\l$, 
and on a self-generated, purely transverse ($\nabla \cdot {\bm A} = 0$), vector potential 
${\bm A}$ 
%
\begin{align}\label{eq:energy_functional_magnetostatic}
{\cal E}[\D, {\bm A},\l] = & E_{\rm var} [\D,\l] 
+ \int_{\rm v} d^3{\bm r} \frac{\left[ \nabla \times {\bm A}({\bm r}) \right]^2 }{2 \mu_0} + \nonumber \\
&
+
\frac{e}{m} \int_{\rm s} d^2 {\bm x}~{\bm p}_{\D}({\bm x}) \cdot {\bm A}({\bm x}) + \nonumber\\ &
+ \frac{e^2}{2 m} \int_{\rm s} d^2{\bm x}~
\rho_{\D}({\bm x}) {\bm A}^2({\bm x})~,
\end{align}
where  $m$ is the bare electron mass in vacuum, $\mu_0$ is the vacuum magnetic permeability, and $\int_{\rm v} d^3{\bm r}$ ($\int_{\rm s} d^2 {\bm x}$) indicates the volume (surface) integral over 
the full space (restricted to cylinder). 
$E_{\rm var}[\D,\l]$ represents the variational electronic energy for the cylindrical geometry. 
We assume $L \gg a$,
so that we can neglect the local curvature of the cylinder and 
treat, for all the practical purposes, 
the electronic system as a 2D system with physical periodic boundary conditions.

Imposing that ${\cal E}$ is stationary with respect to the vector potential, i.e.~$\d {\cal E}/\d {\bm A} = 0$, yields Ampere's law:
\begin{equation}\label{eq:ampere}
- \bm{\nabla}^2 {\bm A} = \mu_0 {\bm J}[{\bm A},\D]~,
\end{equation}
%
where, in cylindrical coordinates ${\bm r} = (r,\theta,y)$, 
the volume current  density reads
\begin{equation}
{\bm J}[{\bm A},\D] = -\frac{e}{m}\d (r- L ) \left[ \bm{p}_{\D}(\bm x) 
+  e \rho_{\D}(\bm x) {\bm A}(\bm r) \right]~.
\end{equation} 
${\bm x} = (L\sin \theta,L \cos \theta ,y)$ represents a point on the  
cylinder surface, and $\rho_{\D}({\bm x})$ and ${\bm p}_{\D}(\bm x)$
are the charge and momentum surface density computed for the 
excitonic state $\ket{\Psi_{\D}}$ and in the 
presence of the vector potential $\bd{A}.$

Solving Eq.~(\ref{eq:ampere}) allows to eliminate 
${\bm A}$ and express the energy ${\cal E}$ as a function of $\D$ only. 
To this extent, we coarse grain the charge and momentum density 
by averaging over the unit cell.
In doing so, we obtain a TRSB state 
with a uniform charge density $\rho_0$ 
and a uniform solenoidal momentum density 
${\bm p}_\D = {\bs{\theta}} P_\D$.
This approximation corresponds to neglecting 
contributions of the flux which are i) perpendicular to the cylinder 
surface and ii) vary over length scales smaller than the lattice parameter.

Ampere's equation is solved by a solenoidal vector potential
${\bm A}(\bm r) = {\bs{\theta}} A_\D \left[ (L/r) \theta_{\rm H}(r-L) + (r/L) \theta_{\rm H} (L-r)\right]$.
Here,  $\theta_{\rm H}$ is the Heaviside step function, 
\replaced{and we have fixed the gauge by requiring the vector potential to vanish}
{which vanishes} for $r/L \to \infty$. 
The solenoidal vector potential $A_\D$ depends 
on the average momentum of the TRSB 
state and on the radius $L$ as following:
\begin{equation}\label{eq:vector_potential_cylinder}
A_\D = - \Phi_0 \frac{P_\D}{h \rho_0} \frac{L}{\ell_0 + L}~.
\end{equation}
Here, $\Phi_0 = h/e$ is the flux quantum and 
$\ell_0 \equiv 2m/(\mu_0 e^2 \rho_0) \simeq 3.5~\mu\mathrm{m}$
is a characteristic length scale of the problem 
set by the electronic density $\rho_0 = 4/a^2$. 
We emphasize that, in Eq.~(\ref{eq:vector_potential_cylinder}), 
$P_\D$ is the average momentum density computed for the 
variational state $\ket{\Psi_{\D}}$, and in the presence of the 
solenoidal vector potential $A_\D,$ so that, eventually,  
$P_{\D}$ and $A_\D$ are simultaneously determined by 
the solution of the variational problem.

For a given $P_\D$, the self-generated 
vector potential leads to a partial cancellation of the 
paramagnetic current that characterizes the TRSB state.
This fact results in a gain of the electronic kinetic energy
which will be balanced by the positive magnetic energy 
\replaced{due to}{per unit volume of} the self-generated flux.
\added{In order to estimate these contributions, we notice that 
the electronic energy is proportional to the cylinder surface, 
whereas the magnetic energy is proportional to the cylinder volume.
We therefore consider the total volume energy 
density by normalizing the total energy over the 
volume of the cylinder $f \equiv {\cal E}/(\pi L^2 L_z) $
Here, ${\cal E} = {\cal E}_{e} + {\cal E}_m$, 
with ${\cal E}_{e/m}$ represent, respectively, 
the electronic/magnetic contributions and 
the limit of an infinitely-long cylinder 
$L_z \to \infty$ is understood.
Denoting with $E^{\rm 2D}_{e} = \frac{{\cal E}_{e}}{2 \pi L L_z}$ the electronic surface energy density, 
and with $E^{\rm 3D}_{m} = \frac{{\cal E}_{m}}{\pi L^2 L_z}$ the magnetic 
volume energy density, the total volume energy density
reads
\begin{equation}
f  = E^{\rm 3D}_{m} + \frac{2 E^{\rm 2D}_{e}}{L}~.
\end{equation}
}

\added{The contribution to the electronic 
surface energy density can be estimated from 
the diamagnetic contribution to the kinetic energy 
$E_{\rm kin} (A) \sim \quave{(P+eA)^2}$, which yields a negative contribution
$\D E^{\rm 2D}_{e} \sim E_{\rm kin} (A) - E_{\rm kin} (A=0) 
\sim 2 e A_\D P_{\D} +  e^2 A_\D^2 \rho_0 
= 
- \frac{P_\D^2}{\rho_0}  \frac{L^2 + 2L\ell_0}{\left( \ell_0 + L \right)^2}$
that increases with the average momentum 
density, \ie moving towards the pure TRSB state at $\varphi=\pi/2$.
At fixed $P_\D$, $\D E^{\rm 2D}_{e}$ increases sub-linearly with 
$L/\ell_0$, saturating for $L \gg \ell_0$,
so that the electronic gain in the  volume energy density scales as 
${\D E^{2d}_{e}}/ L \sim - P_{\D}^2 \frac{\ell_0}{L}$.
On the contrary, by computing the flux density $B = 2 A_\D/L$, it is easy 
to check that the magnetic energy per unit volume 
$E_{m}^{\rm 3D} = B^2/2 \mu_0$ decreases, for $L \gg \ell_0$, as 
$\D E_{m}^{\rm 3D} \sim P_\D^2(\ell_0/L)^2$.
This indicates that, by increasing the cylinder radius, 
the energy loss of the TRSB  decreases faster than the energy gain.
As a result,  we expect the overall energy gain of the TRSB 
due to the sourcing of the magnetic flux to be negligible for 
$L \ll \ell_0$ and to become sizeable in the opposite limit.}

\deleted{Specifically, the diamagnetic contribution to the \added{electronic} kinetic 
energy $E_{\rm kin} (A) \propto \quave{(P+eA)^2}$, yields a negative 
contribution $\D E_{\rm kin}(A) = E_{\rm kin} (A) - E_{\rm kin} (A=0) 
= 2 e A_\D P_{\D} +  e^2 A_\D^2 \rho_0 
$ $= 
- \frac{P_\D^2}{\rho_0}  \frac{L^2 + 2L\ell_0}{\left( \ell_0 + L \right)^2}$
which increases with the average momentum 
density, \ie moving towards the pure TRSB state at $\varphi=\pi/2$. 
At the same time, for fixed $P_\D$, the energy gain in the 
electronic energy  
increases with $L/\ell_0$.
Moreover, as 
it is clear from Eq.~\eqref{eq:vector_potential_cylinder}, 
 the magnetic energy per unit volume 
 decreases with 
$L/\ell_0$.
As a result, 
we expect the overall energy gain of the TRSB due 
to the sourcing of the magnetic flux to be negligible for 
$L \ll \ell_0$ and to become sizeable in the opposite limit.}


We confirm the above expectations by 
solving the full variational problem,
namely by self-consistently solving  Eq.~(\ref{eq:vector_potential_cylinder}),
together with the optimization of the Lagrange parameter and 
the variational wave-function 
$\ket{\Psi_{\D}}$~\cite{suppl}. 
\deleted{We obtain the 
total volume energy density $f(\D) \equiv {\cal E}/(\pi L^2 L_z)$ as a function of the 
order parameter, where the limit of an infinitely-long cylinder $L_z \to \infty$ is 
understood.}
In Fig.~\ref{fig:fig3}(a) we 
\replaced{compare the total 
volume energy densities at fixed 
cylinder radii for several phases moving 
from the pure CDSB to the pure TRSB.
}{plot the total energy density for increasing 
cylinder radius.} 
The energy of the pure CDSB state does not depend 
on~$L$. On the contrary, 
the TRSB states experience an energy lowering which increases as a function of $L$. 
As the radius of the solenoid increases, the energy minimum 
shifts from the CDSB state to the TRSB state for 
$L>L_\ast$, with a critical radius $L_\ast  \approx 40~\mu \mathrm{m}.$
In the TRSB state, the solenoid is characterized by a macroscopic 
magnetic moment sustained by a ground-state persistent 
solenoidal current, which represents an alternative 
manifestation of the spontaneous breaking of the $x \to -x$
crystal symmetry.
\added{In Fig.~\ref{fig:fig3}(b) we show that an external flux 
can further lower the energy barrier between the CDSB and  TRSB states, 
thus highlighting the possibility of tuning the critical radius $L_\ast$.}

\replaced{We now comment on the limit of a large solenoid. 
In the limit $L/\ell_0 \to \infty$, the diamagnetic current perfectly 
cancels the paramagnetic one leading to a TRSB state with 
vanishing surface current density.   
We emphasize that, for any macroscopic value of $L/\ell_0$, 
the cylinder remains topologically distinct from an infinite 2D sheet. 
Such a distinction between the two geometries can be traced back
to the fact that the TRSB is stabilized by self-generated magnetic flux
which carries information about the geometry. 
Indeed, the solution of Ampere's equation in the 
cylindrical geometry enforces the physical vector 
potential to vanish for $r/L \to \infty$, whereas, 
for an infinite 2D sheet, the vector potential diverges for 
$r/L \to \infty$~\cite{suppl}. 
In particular, at variance with the cylinder case, 
for an infinite 2D sheet the magnetic energy associated with 
the sourced flux would be infinite. 
As a result, no energy gain for the TRSB would be possible
and the CDSB state remains the only stable state,
in accordance to the aforementioned no-go theorems.
}{
We notice that for $L \gg \ell_0$ the diamagnetic 
contribution leads to a perfect cancellation of 
the paramagnetic current density, and, 
therefore to a vanishing contribution of the positive 
magnetic energy per unit volume.
We stress the fact, due to the physical boundary 
conditions, the physical vector potential vanishes for $r/L \to \infty$, 
thus making the system topologically distinct from an infinite 2D sheet 
for any macroscopic value of $L/\ell_0$.}

\deleted{Finally, in Fig.~\ref{fig:fig3}(b) we show that an external flux 
can further lower the energy barrier between the CDSB and  TRSB states, 
thus highlighting the possibility of tuning the critical radius $L_\ast$.}


\section{{Discussion and} Conclusions.}
We have shown a controlled path to the stabilization
of a spontaneous TRSB phase of purely orbital 
character issuing from an excitonic instability.
The TRSB state is achieved by topological constraints 
on the system which impose physical boundary conditions and 
enable the direct coupling  between the excitonic order parameter 
and a self-generated magnetostatic potential that cannot be gauged away.
The TRSB excitonic phase with self-generated vector 
potential is equivalent to the superradiant excitonic 
insulating phase discussed in Ref.~\cite{giacomo_antoine_sxi}  
once the no-go theorems about {spatially-uniform} vector potentials are 
properly taken into account~\cite{andolina_nogo,andolina_epjp_2022,nataf_prl_2019,andolina_prb_2020,guerci_prl_2020}.

Our results have direct implications on 
the disentangling of coupled excitonic and structural 
transitions.
At variance with the CDSB state, the TRSB excitonic state is charge 
symmetric and is expected to have no direct coupling with lattice distortions.  
Estimating the relative energy gain between the lattice distortion 
and the flux generation discussed here is beyond the scope 
of this work.
Nonetheless, the control of the energy barrier 
between the TRSB and CDSB states with an 
external flux---see Fig.~\ref{fig:fig3}---highlights the 
intriguing possibility of a magnetic  tuning of coupled 
structural and excitonic phase transition.
As such, our results suggest an equilibrium route for the disentangling 
of excitonic and structural phase transitions.
\replaced{For example, materials of interest include the 
above mentioned Ta$_2$NiSe$_5$ for which 
attempts to disentangle the two transitions 
focused so far only on} 
{which is alternative to} the use of time-dependent probes~\cite{golez_prb2022,yann_gallais_tns,averitt_tns}.
\replaced{We emphasize that the specific geometry 
for the stabilization of the TRSB phase depends on the 
precise symmetry of the order parameter and may vary from case to case. 
Specific material realizations and the interplay with 
lattice distortions represent 
natural directions for future research.}{The specific geometrical and material realizations 
represent interesting future research directions.}
This may include, for example, the interplay 
between geometrical constraints discussed here 
and the structured electromagnetic vacuum inside 
chiral cavities~\cite{hubener_naturematerials_2021}.
\added{Moreover, our results can also have implications for other types of intertwined charge and chiral orders 
as discussed, for example, in the case of kagome 
metals~\cite{chunyo_guo_kagome,grandi_kagome}}.

The experimental discovery of excitonic TRSB phases would 
provide us with an entirely new family of quantum systems displaying 
non-trivial time-reversal symmetry breaking of orbital origin, together 
with e.g. Chern insulators (such as those recently discovered in twisted bilayer graphene~\cite{serlin_science_2019,nuckolls_nature_2020,
xie_nature_2021,stepanov_prl_2021,tschirhart_science_2021} and 
ABC-trilayer graphene~\cite{chen_nature_2020}) and 
chiral superconductors (see, for example, 
Refs.~\onlinecite{volovik_oup_2003,xia_prl_2006,qi_rmp_2011,
kallin_repprogphys_2016,bernevig_princeton_2013} and references therein).

\section*{Acknowledgements}
This work was funded by the Swiss National Science Foundation 
through an AMBIZIONE grant (\#PZ00P2\_186145) and by 
the MUR - Italian Minister of University and Research 
under the ``Rita Levi-Montalcini" program (G.M.).
M.P. is supported by the European Union's Horizon 2020 research and innovation programme under the grant agreement No.~881603 - GrapheneCore3 
{and the Marie Sklodowska-Curie grant agreement No.~873028}, 
by the University of Pisa under the ``PRA - Progetti di Ricerca di Ateneo'' (Institutional Research Grants) -  Project No.~PRA\_2020-2021\_92 
``Quantum Computing, Technologies and Applications'', 
and by the MUR - Italian Minister of University and Research under 
the ``Research projects of relevant national interest  - PRIN 2020''  - 
Project No.~2020JLZ52N, 
title ``Light-matter interactions and the collective behavior of quantum 2D materials (q-LIMA)''. 

It is a great pleasure to thank Antoine Georges, Andrew Millis, and Aharon Kapitulnik for inspiring discussions.

\end{document}